\documentclass[a4paper]{article}
\usepackage{graphicx,amssymb}

\begin{document}
\title{An ultra sensitive radio frequency single electron transistor working up to 4.2 K}

\author{\small{Henrik Brenning, Sergey Kafanov, Tim Duty, Sergey Kubatkin and Per Delsing} \and{ \small{ \emph{ Quantum Device Physics Laboratory, Chalmers University of
Technology}}}  \and{ \small{ \emph{S-412 96 Gothenburg, Sweden}}}}

\maketitle

\begin{abstract}
We present the fabrication and measurement of a radio frequency
single electron transistor (rf-SET), that displays a very high
charge sensitivity of $1.9$ $\mu$e/$\sqrt{Hz}$ at $4.2$ K. At $40$
mK, the charge sensitivity is $0.9$ and $1.0$ $\mu$e/$\sqrt{Hz}$
in the superconducting and normal state respectively.  The
sensitivity was measured as a function of radio frequency
amplitude at three different temperatures; $40$ mK, $1.8$ K and
$4.2$ K.
\end{abstract}

\newpage

The radio frequency single electron transistor (rf-SET) is the
most sensitive detector of charge to date. Unlike the conventional
single electron transistor (SET), it is not bandwidth limited by
the resistance-capacitance product of the SET resistance and the
parasitic lead capacitance.  Typically the rf-SET displays high
bandwidth, $10$ MHz, in combination with a charge resolution of
the order of $10^{-5}$ e/$\sqrt{Hz}$.  Although the conventional
SET is theoretically more sensitive than the rf-SET
\cite{korotkov}, the rf-SET can operate at frequencies where 1/f
noise is negligible, which makes the rf-SET more sensitive in
experimental conditions. The improved band width and charge
sensitivity have made the rf-SET a good choice when measuring
solid state charge qubits\cite{qubit1} \cite{qubit2}, charging of
quantum dots \cite{dot1} and single electron transport
\cite{electron_count2} \cite{electron_count1}. In rf measurements
of the SET, the impedance of the SET is matched to the
characteristic impedance of a co-axial cable by a resonance
circuit. The difficulty of making a tank circuit with a high
Q-value as well as a high operating frequency makes rf measurement
of a SET practical only when the SET is relatively low ohmic.
Therefore, SET resistances in the range $20$ to $200$ k$\Omega$
are desirable. As the tunnel junctions are made smaller the
charging energy (E$_{C}$) increases, which increases both the
charge sensitivity of the SET and the maximum operating
temperature. However, since the tunnel resistance is inversely
proportional to the junction area, this also increases the
resistance, and eventually the resistance becomes too large for
rf-SET operation. With conventional aluminum angle evaporation, it
has been difficult to make tunnel junctions smaller than $100*100$
nm$^{2}$ (E$_{C} \approx 1K$) without increasing the device
resistance too much. Hence, the operating temperature range of the
rf-SET has been limited to roughly a few hundered mK. By using low
oxidation pressure \cite{fab1} we here show that it is possible to
combine high E$_{C}$ with rf operation.  There are numerous
experiments that require the bandwidth and sensitivity of the
rf-SET, and are therefore performed at millikelvin temperatures.
With a higher operating temperature of the rf-SET, many of these
experiments could also be conducted at $4.2$ K. Other experiments
now use conventional SETs with a higher charging energy, and also
resistance, to enable measurements at higher temperatures.  By
switching to a rf-SET operating at $4.2$ K some of these
experiments, such as electron counting \cite{electron_count1} and
the scanning-SET \cite{scanSET}, could gain in sensitivity and
bandwidth.

In this letter, we describe the measurement of a rf-SET working
from $40$ mK to $4.2$K of the order of $1$ $\mu$e/$\sqrt{Hz}$

The SET was fabricated with 2-angle evaporation of aluminum on
SiO$_{2}$ and in situ oxidation.  The details are described in
\cite{fab1}. In the fabrication a very low oxidation pressure was
used, which resulted in very thin tunnel barriers.  Since the
tunnel resistance depends exponentially on the barrier thickness,
this improved the specific conductance of the barriers without
increasing the specific capacitance. The data presented here are
taken on a device which had an asymptotic serial tunnel resistance
R$_{\Sigma}$ = $25$ k$\Omega$, in spite of its very small size
(see figure 1). The relatively low resistance made strong
tunneling contributions sizable. The effect of this was two fold.
First, the Coulomb diamonds were smeared due to strong tunneling,
even at low temperature, which made it difficult to fit asymptotes
to the Coulomb diamond edges and hence to determine the charging
energy (E$_{C}$). Second, the nominal E$_{C}$ (as determined of
the total island capacitance) of the SET was lowered to an
effective E$_{C}$ \cite{grabert1}. The charging energy of the SET
was estimated by fitting asymptotes to the coulomb diamond (see
figure 3b) and the resulting charging energy was E$_{C}$=$18 \pm
2$ K, which corresponds to a total island capacitance C$_{\Sigma}$
= $58$ aF. One junction capacitance was slightly larger than the
other, $33$ aF compared to $25$ aF, which indicates that the
geometrical symmetry of the SET was good.

The experimental SET up is depicted in fig. 1. Various filters are
not shown to simplify the figure. The rf signal is transmitted
from room temperature via a directional coupler at $4.2$ K, and is
reflected at the combined SET/tank circuit.  At resonance, the
reflected power depends on the resistance of the SET and the tank
circuit parameters, \textit{i.e.} P$_{R}$=P$_{A}$$\left( 1 - 4
Q^{2} \frac{Z_{0}}{R_{d}} \right)$ \cite{wahlgren}, where P$_{R}$,
P$_{A}$, Q are the reflected rf power, the applied rf power and
the Q-value of the tank circuit.  Z$_{0}$ and R$_{d}$ denote the
characteristic impedance of coaxial cable connected to the tank
circuit and the resistance of the SET respectively.  In our set
up, the Q-value was approximately $11.6$ and the characteristic
impedance was $50\Omega$. To measure the charge sensitivity of the
SET, we proceeded as in \cite{abdel}.  The SET was excited with a
$1.5$ MHz gate voltage which amplitude modulates the carrier
frequency ($345$ MHz) and produces sidebands in the frequency
spectrum of the reflected rf signal. The sidebands and main
frequency can be seen in the upper inset in figure 2. The
sensitivity of the SET is then calculated by comparing the height
of the side band peak with the noise floor, \textit{i.e.} the
signal to noise ratio.  The sensitivity $\delta$Q, is:
\begin{equation}
\delta Q=\frac{\Delta
q_{rms}}{\left(\sqrt{2B}*10^{\frac{SNR}{20}}\right)}
\label{correct}
\end{equation}
Here, $\Delta$q$_ {rms}$ is the applied root mean square gate
charge, B is the resolution bandwidth and SNR is the signal to
noise ratio in dB.  The additional factor $\sqrt{2}$ in the
denominator as compared to \cite{abdel}, includes the
contributions from both sidebands since information can be
extracted by homodyne mixing from both sidebands. We measured the
sensitivity for different rf amplitudes, and for each rf amplitude
we varied V$_{SD}$ and V$_{g}$ to find the optimum bias point.
This procedure was repeated at the temperatures $4.2$, $1.8$ and
$40$ mK.

At $4.2$ K, the best SNR was $22.88$ dB, $\Delta$q$_ {rms}$
$0.0044$ e$_ {rms}$ and B was $15$ kHz, which results in a
sensitivity of $1.9 \pm 0.1$ $\mu$e/$\sqrt{Hz}$. The current
voltage characteristics shown in figure 2 display a large
modulation of approximately $20$ nA of the source drain current
(I$_{SD}$) with respect to the gate voltage (V$_{g}$), despite the
relatively high temperature. It is clear that rf operation of the
SET should be possible. In the lower inset of figure 2, the
reflected power is plotted as a function of V$_{g}$ and V$_{SD}$,
where the highest signal to noise ratio is achieved close to zero
bias.  A closer inspection shows that this maximum was achieved
with V$_{SD}$=$-0.05$ mV, \textit{i.e.} near a pure rf mode
\cite{abdel} measurement. The optimum sensitivity in the pure rf
mode has been calculated by Korotkov and Paalanen \cite{korotkov}:
\begin{equation}
\delta Q = 2.65 e \left( R_{\Sigma} C_{\Sigma} \right)^{1/2}
\left( k_{B} T C_{\Sigma} / e^{2} \right)^{1/2}
\end{equation}
where  k$_{B}$ and T stand for the Boltzman constant and absolute
temperature.   If the total capacitance and resistance of the
measured SET is used, this formula results in a maximum
theoretical sensitivity of $1.2$ $\mu$e/$\sqrt{Hz}$ at $4.2$ K.
The charge sensitivity is therefore approximately $1.6$ times
worse than the theoretical minimum.

At $40$ mK (see figure 3), the sensitivity improved approximately
by factor of two. In the superconducting case, the sensitivity was
$0.9 \mu \pm 0.1 \mu$e$\sqrt{Hz}$.  Several factors contributes to
the uncertainty.  The spectrum analyzer has an accuracy better
than $0.01$ dB, and calibrating the voltage necessary to induce
$1$ e$_ {rms}$ on the gate has an uncertainty of approximately $4$
\%. In addition to these systematic errors, the gate bias points
can vary due to fluctuating charges in the vicinity of the SET.
Two consecutive measurements separated by $24$ hours resulted in
two nearly equal maximum sensitivities ($0.85$ and $0.88$
$\mu$e/$\sqrt{Hz}$ in the superconducting state). The combined
uncertainty is $\sim 7$ \%. The sensitivity in the normal case was
$1.0 \pm 0.1$ $\mu$e/$\sqrt{Hz}$, and in both the superconducting
and the normal state case the V$_{SD}$ was small at the optimum
bias point; $0.1$ mV. At this temperature, however, the
theoretical maximum sensitivity is roughly five times better than
what we measured. A plot of the shot noise of the SET as a
function of current (see figure 3a) with the SET at $40$ mK and in
the normal state, shows that the noise temperature (T$_{n}$) of
the amplifier clearly contributes.  T$_{n}$ is approximately $10$
K referred to the tank circuit, \textit{i.e.} substantially higher
than the nominal noise temperature of the amplifier alone, which
is $2$ K. In the charge sensitivity measurements we used a
resolution bandwidth of $15$ kHz, which translates the noise
temperature to a $-92$ dBm noise floor.  This level co-insides
with the noise floor of the $40$ mK sensitivity measurements,
\textit{i.e.} the sensitivity is degraded by the amplifier noise.
Self heating of the SET, which grows larger with a smaller SET,
could also limit the performance of the SET in the $40$ mK
measurements. Since the applied voltage is large and the island is
very small (see figure 1), the electron temperature may well be of
the order of $1$ K.

In figure 4, we see how the charge sensitivity depends on the
applied rf signal for the measurements at $40$ mK, $1.8$ K and
$4.2$ K.  For each of these rf amplitudes, V$_{SD}$ and V$_{g}$
has been optimized to find the best sensitivity.  As seen in
figure 4, the maximum sensitivity is found at rf amplitudes
$-82.5$ dBm ($40$ mK, superconducting state), $-77.5$ dBm ($40$
mK, normal state), $-80.5$ dBm ($1.8$ K, normal state) and $-72.5$
dBm ($4.2$ K).

The SET reported here had a $5-6$ times larger E$_{C}$ compared to
the sample in \cite{abdel}, but lower tunnel resistance.  As shown
by Korotkov and Paalanen, the size of the optimum rf signal
increases with increasing E$_{C}$ \cite{korotkov}. Therefore, we
could apply a larger rf signal and still cover a part of the
coulomb diamond where dI$_{SD}$/dV$_{g}$ was large. The applied rf
signal was between $12.5$ (at $40$ mK, superconducting state) and
$17.5$ ($4.2$ K) dB larger than in \cite{abdel}. This yielded a
better signal to noise ratio, even when the higher noise
temperature of the amplifier compared to \cite{abdel} has been
taken into account, and hence a better charge sensitivity. Using
the modified sensitivity formula (\ref{correct}), the sensitivity
of the previously best reported result \cite{abdel} is $2.3 $
$\mu$e/$\sqrt{Hz}$, which should be compared to our sensitivities:
$1.9$ $\mu$e/$\sqrt{Hz}$ (at $4.2$ K), $1.0$ $\mu$e/$\sqrt{Hz}$
($40$ mK, normal state SET) and $0.9$ $\mu$e/$\sqrt{Hz}$ ($40$ mK,
superconducting SET).

In summary, we have measured a charge sensitivity for a rf-SET
that is better than the previously best reported value both at
$40$ mK, and at $4.2$ K.  This is due to high charging energy and
low tunnel resistance. The higher operating temperature of this
device makes it possible to perform rf-SET measurements at $4.2$ K
rather than at mK temperatures.

We were supported by the Swedish VR and SSF, and by the Wallenberg
foundation.

\newpage
\Large{Caption figure 1} \newline $\quad$
\newline $\quad$ \normalsize{The schematics of the rf-measuremtent.
Filters are not shown.  The inset in the top right corner shows an
scanning electron microscope image of the SET, with the exception
of the gate electrode.  ``I'',``S'' and ``D'' stand for island,
source and drain respectively.  The scale bar is $100$ nm}
\newline $\quad$
\newline $\quad$
\Large{Caption figure 2} \newline $\quad$
\newline $\quad$ \normalsize{(Color online).  Measurements at
$4.2$ K.  The current-voltage characteristics, I$_{SD}$ as a
function of V$_{SD}$ for various V$_{g}$ voltages. In the upper
left inset, the reflected power (RP) as a function of frequency is
shown. The two sidebands are situated $1.5$ MHz to the left and
right of the main frequency. The lower right inset is a color plot
of the signal to noise ratio of the side bands as a function of
V$_{SD}$ and V$_{g}$.}
\newline $\quad$
\newline $\quad$
\Large{Caption figure 3} \newline $\quad$
\newline $\quad$
 \normalsize{(Color online). Measurements at $40$ mK with the SET in the superconducting state.
 a). The current-voltage characteristics, for various
 V$_{g}$. In the upper left inset, the shot noise of the SET as a
 function of I$_{SD}$ where the noise is collected in a $8$ MHz span at the output of the cold amplifiers.  The two asymptotes intersect at
 the amplifier noise contribution, which is estimated to $10$ K.  In the lower right inset the
 signal to noise ratio of the sideband, measured in dB, is plotted as a function of V$_{SD}$ and V$_{g}$. The mark ``x'' shows
 the optimum bias point for P$_{i}$=$-102.5$ dBm, and the dotted circle marks the optimum bias point for P$_{i}$= $-82.5$dBm.  P$_{i}$
 signifies the power of the incident rf signal at the tank circuit.  b) The reflected power of the tank circuit as a function
 of V$_{SD}$ and V$_{g}$.  The dotted red lines show a fit to coulomb blockade diamond, with E$_{C}$=$18 \pm 2 K$}

\newpage
 \Large{Caption figure 4} \newline $\quad$
\newline $\quad$
 \normalsize{(Color online).The sensitivities as a function of the applied rf power at the
 tank circuit for three different temperatures.}

\newpage

\Large{Figure 1, author H. Brenning}
\newline $\quad$
\newline $\quad$
\newline $\quad$
\begin{figure}[!thp]
\begin{center}
  \includegraphics[bb=0 0 361 155, width=8.5cm]{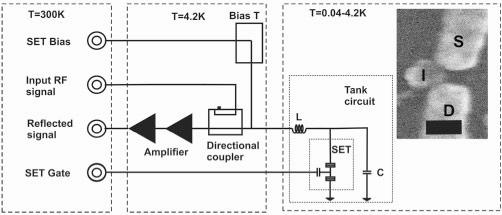}\label{tank}\\

\end{center}
\end{figure}

\newpage
\Large{Figure 2, author H. Brenning}
\newline $\quad$
\newline $\quad$
\newline $\quad$
\newline $\quad$
\begin{figure}[!thp]
\begin{center}
  \includegraphics[bb=0 0 361 318, width=8.5cm]{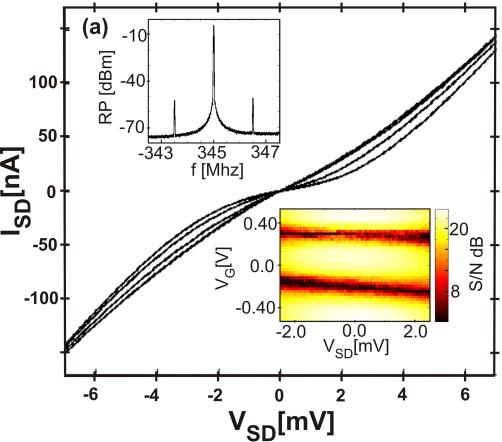}\label{db4p2}\\

\end{center}
\end{figure}

\newpage
\Large{Figure 3, author H. Brenning}
\newline $\quad$
\newline $\quad$
\newline $\quad$
\newline $\quad$
\begin{figure}[!thp]
\begin{center}
  \includegraphics[bb=0 0 361 462, width=8.5cm]{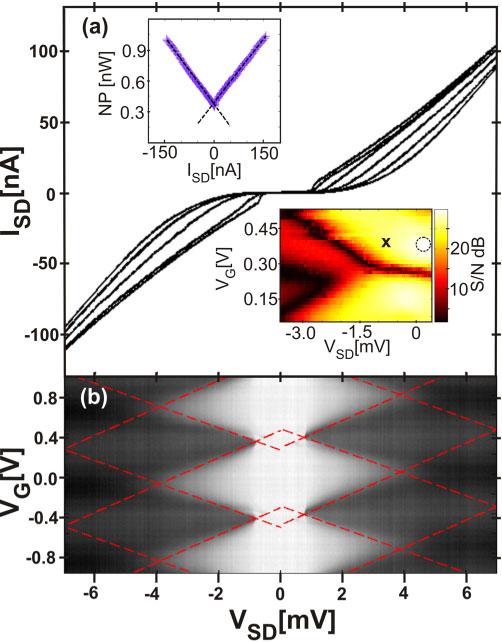}\label{db90mk}\\

\end{center}
\end{figure}

\newpage
\Large{Figure 4, author H. Brenning}
\newline $\quad$
\newline $\quad$
\newline $\quad$
\newline $\quad$
\begin{figure}[!thp]
\begin{center}
  \includegraphics[bb=0 0 361 269, width=8.5cm]{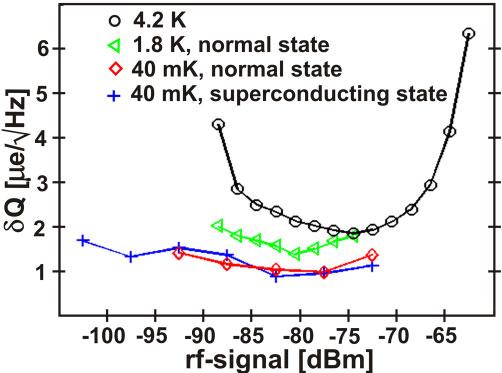}\label{sens3}\\

\end{center}
\end{figure}

\end{document}